\documentclass[aps, pra, twocolumn, superscriptaddress, amsmath,  tightenlines, longbibliography]{revtex4-2}

\usepackage{tikz}
\usepackage{graphicx} % needed for figures
\usepackage{epstopdf}
\usepackage{dcolumn}
\usepackage{graphicx}
\usepackage{mathrsfs}
\usepackage{subfigure}
\usepackage{booktabs}
\usepackage{amsmath}
\usepackage{physics}
\usepackage{dsfont}
\usepackage{amstext}
\usepackage{amssymb}
\usepackage{amsbsy}
\usepackage{bbm}
\usepackage{amsthm}
\usepackage{color}
\usepackage[colorlinks,citecolor=blue]{hyperref}

\usepackage{url}
\usepackage[colorlinks]{hyperref}
\hypersetup{%
	plainpages=true,
	breaklinks=true,       %not default in dvips mode, so we must specify
	hypertexnames=false,  %not ideal, but needed when pagenums duplicate (`i' vs. `1')
	pageanchor=true,
	colorlinks=true,
	linkcolor={blue},
	citecolor={red},
	urlcolor={blue},
	anchorcolor={black}
}

 \makeatletter

\newcommand{\Rmnum}[1]{\expandafter\@slowromancap\romannumeral #1@}
\makeatother

\begin{document}

% Be sure to use the \title, \author, \affiliation, and \abstract macros
% to format your title page.  Don't use lower-level macros to  manually
% adjust the fonts and centering.

\title{Searching Axion-like Dark Matter by Amplifying Weak Magnetic Field \\
with Quantum Zeno effect}
% In a long title you can use \\ to force a line break at a certain location.
	\author{Jing Dong}
    %\thanks{These authors contributed equally to this work.}
\affiliation{School of Physics and Astronomy, Applied Optics Beijing Area Major Laboratory, Beijing Normal University, Beijing 100875, China}
\affiliation{ Institute of Physics, Chinese Academy of Sciences, Beijing 100190, China}
\affiliation{ University of Chinese Academy of Sciences, Beijing 100049, China}
\affiliation{Key Laboratory of Multiscale Spin Physics, Ministry of Education, Beijing Normal University, Beijing, China}
 \author{W. T. He}
\affiliation{Quantum Dynamics Unit, Okinawa Institute of Science and Technology, Tancha 1919-1, Okinawa 904-0495, Japan}
\author{S.-D. Zou}
	%\altaffiliation{Department of Physics, Applied Optics Beijing Area Major Laboratory, Beijing Normal University, Beijing 100875, China}%Lines break automatically or can be forced with \\
\affiliation{School of Physics and Astronomy, Applied Optics Beijing Area Major Laboratory, Beijing Normal University, Beijing 100875, China}
    \author{D. L. Zhou}
\affiliation{ Institute of Physics, Chinese Academy of Sciences, Beijing 100190, China}
\affiliation{ University of Chinese Academy of Sciences, Beijing 100049, China} 
	\author{Qing Ai}
	\email{aiqing@bnu.edu.cn}
	%\altaffiliation{Department of Physics, Applied Optics Beijing Area Major Laboratory, Beijing Normal University, Beijing 100875, China}%Lines break automatically or can be forced with \\
\affiliation{School of Physics and Astronomy, Applied Optics Beijing Area Major Laboratory, Beijing Normal University, Beijing 100875, China}
\affiliation{Key Laboratory of Multiscale Spin Physics, Ministry of Education, Beijing Normal University, Beijing, China}

%When submitting the manuscript for review, do not include the author's name or institution
%\author{Daniel V. Schroeder}
%\email{dschroeder@weber.edu} % optional
%\altaffiliation[permanent address: ]{101 Main Street, Anytown, USA} % optional second address
% If there were a second author at the same address, we would put another
% \author{} statement here.  Don't combine multiple authors in a single
% \author statement.
%\affiliation{Department of Physics, Weber State University, Ogden, UT 84408-2508}
% Please provide a full mailing address here.

%\author{David P. Jackson}
%\email{ajp@dickinson.edu}
%\affiliation{Department of Physics, Dickinson College, Carlisle, PA 17013}

% See the REVTeX documentation for more examples of author and affiliation lists.

\date{\today}

\begin{abstract}
The enhancement of weak signals and the detection of hypothetical particles, facilitated by quantum amplification, are crucial for advancing fundamental physics and its practical applications.
Recently, it was experimentally observed that magnetic field can be amplified by using nuclear spins under Markovian noise, [\href{https://link.aps.org/doi/10.1103/PhysRevLett.133.191801}{H. Su, \textit{et al}., Phys. Rev. Lett. \textbf{133}, 191801 (2024)}]. Here, we theoretically propose amplifying the magnetic-field signal by using nuclear spins by the quantum Zeno effect (QZE). Under identical conditions, we demonstrate that compared to the Markovian case the amplification of the weak magnetic field can be enhanced by a factor about $e^{1/2}$ under a Gaussian noise.
Moreover, through numerical simulations we determine the optimal experimental parameters for amplification conditions. This work shows that the combination of the QZE and spin amplification effectively enhances the amplification of the weak magnetic field. Our findings may provide valuable guidance for the design of experiments on establishing new constraints of dark matter and exotic interactions in the near future.
\end{abstract}
% AJP requires an abstract for all regular article submissions.
% Abstracts are optional for submissions to the "Notes and Discussions" section.

\maketitle % title page is now complete

\section{Introduction} % Section titles are automatically converted to all-caps.
% Section numbering is automatic.
According to astrophysical observations, roughly {five sixths} of the matter in the universe
remains dark \cite{David2015science}. However, direct detection of its interactions with particles and fields of the standard model remains elusive \cite{Bertone2018RevModPhys}. There are a variety of particle candidates for dark matter, such as quantum chromodynamics axion and axion-like particles (ALPs) \cite{Sikivie2021RevModPhys}, new $Z'$ bosons \cite{Andreev2022PRL}, new spin-1 bosons \cite{Su2021science} and dark photons \cite{Hochberg2019PRL,Manley2021PRL}. Axions are prominent dark-matter and dark-energy candidates, which are introduced as a compelling solution to the strong-CP problem beyond the standard model. However, it is insufficient to search for ALPs by traditional particle-physics techniques with light quanta such as the Large Hadron Collider \cite{David2017science,Safronova2018RMP}. Therefore, experimental searches for axion-like dark matter are based on their non-gravitational interactions with particles and fields of the standard model \cite{David2017science,Safronova2018RMP,Bradley2003RMP,Anastassopoulos2017,Braine2020,Ouellet2019PRL,Jiang2021SCIENCE}.

%Recently, a variety of works have focused their searches on axion¨Cnucleon interactions

As known to all, quantum amplification plays an important role in quantum metrology and finds applications in the measurements of weak field and force \cite{Kotler2011nature,Burd2019Science,Jiang2022PRL,Schaffry2011PRL}, and optical amplification \cite{Zavatta2011Nature} to search for new physics beyond the standard model \cite{Bradley2003RMP}. Besides, the magnetic-field amplification finds applications in a wide range of searching axion-nucleon interactions, such as ALPs and axion bursts from astrophysical events \cite{Afach2021NATURE,Dailey2021nature}. Under the state-of-art experiments, the sensitivity has exceeded astrophysical limits \cite{Chang2018} by several orders of magnitude. We note that both electron and nuclear spins have shown great potential for realizing signal amplification. For example, the overlapping spin ensemble, e.g. $^{129}$Xe-$^{87}$Rb, is used in self-compensating comagnetometers \cite{Almasi2020PRL,Lee2018PRL,Vasilakis2009PRL}. Recently, the magnetic-field amplification using $^{129}$Xe noble gas overlapping with spin-polarized $^{87}$Rb atomic gas is demonstrated, which has achieved a significant improvement in amplification of weak-field measurements under the Markovian noise, i.e., a constant spin relaxation rate \cite{Jiang2024PNAS}.

On the other hand, the QZE describes the phenomenon that a quantum system's dynamic evolution drastically slows down when measured frequently enough \cite{Misra1977,Peres1990PRA,Block1991PRA,Aiqing2010PRA,Xu2011PRA}. The QZE plays a pivotal role in various domains of quantum science. One of the applications of the QZE is the quantum measurements, which can suppress the detrimental effects of decoherence \cite{Facchi2008,Wu2004PRA}. By performing frequent error-correction measurements, the QZE helps stabilize the quantum state of qubits, thereby prolonging their coherence time and enhancing the reliability of quantum computations. Both theoretically and experimentally, it has been demonstrated that the QZE can enhance the quantum metrology by using the maximum-entangled states {\cite{Chin2012PRL,Matsuzaki2011PRA,Long2022PRL}}. So far, the QZE has been experimentally observed in a number of physical systems such as trapped ions \cite{Itano1990PRA,Balzer2000OC}, ultracold atoms \cite{Fischer2001PRL,Wilkinson1997nature,Streed2006PRL}, molecules \cite{Nagels1997PRL}, Bose-Einstein condensates \cite{Schafer2014Nature}, nitrogen-vacancy centers \cite{Kalb2016Nat} and superconducting quantum circuits  \cite{Potocnik2018Nat,Kakuyanagi2015NJP,Slichter2016NJP,Bretheau2015Science}.

To achieve the more significant amplification of external magnetic fields, in this paper, we propose a magnetic-field amplification using noble gas overlapping with spin-polarized alkali-metal gas by the QZE, which can achieve an improvement in the amplification of weak-field measurements as compared to the Markovian case. In weak magnetic fields, we obtain the analytical solution to the dynamics of $^{129}$Xe spins and amplification function, which present an enhancement of $e^{1/2}$ compared to Markovian case. In the case of strong magnetic fields, we further examine the response of the $^{129}$Xe spins by numerical simulations under various conditions.
Then, we provide the optimal measurement parameters for the practical experiments, i.e., the detuning between the Larmor frequency and the frequency of magnetic field, the decoherence time, and the magnitude of the magnetic field. We anticipate that the present amplification technique could stimulate possible
applications in applied and fundamental physics.

The rest of the paper is organized as follows. The model is set up and the response of polarized $^{129}$Xe spins to a transverse oscillating magnetic field is derived in Sec.~\ref{sec:model}. Then,
the response of polarized $^{129}$Xe spins in the case of weak magnetic fields is investigated in Sec.~\ref{sec:linear}. In Sec.~\ref{sec:nonlinear}, the transverse polarization of the polarized $^{129}$Xe spins is studied and the optimal amplification is explored in the case of strong magnetic fields{.}
A summary is {concluded} in Sec.~\ref{sec:Conclusion}.

\section{model}
\label{sec:model}

According to Ref.~\cite{Su2024PRL}, there are two cells containing nuclear spins, i.e., the source and the sensor cells. In order to polarize the {Rb} spins and thus the $^{129}$Xe nuclear spins, a static magnetic field with strength $B_0$ is applied along $z$-axis. Due to the axion-mediated interaction between polarized neutrons, 
the $^{129}$Xe nuclear spins are also subject to a time-dependent transverse magnetic field. As a result,
the $^{129}$Xe nuclear spin in a magnetic field is described by the following Hamiltonian
\begin{equation}
\hat{H}=-\gamma_n\mathbf{B}(t)\cdot{\hat{\mathbf{I}}},
\end{equation}
where $\mathbf{B}(t)=B_0 \hat{z} + B_{\text{ac}} \cos(2\pi \nu t) \hat{y}$ is the total field experienced by $^{129}$Xe nuclear spins with $\nu$ being the Larmor frequency of the nuclear spins in the source cell, $\gamma_n = 2\pi \times 11.78$~Hz/$\mu$T \cite{Jiang2024PNAS} and $\hat{\mathbf{I}}=(\hat{I}_x,\hat{I}_y,\hat{I}_z)$ are the gyromagnetic ratio and the angular momentum operators of $^{129}$Xe, respectively. Thus, the Larmor frequency of $^{129}$Xe is 
\begin{align}
\nu_0=\frac{\gamma_n B_0}{2\pi},\label{eq:Larmor}
\end{align}
{where $\gamma_n$ is the gyromagnetic ratio of $^{129}$Xe.} The spin Hamiltonian with the total field reads explicitly
\begin{align}
\hat{H} &=-\gamma_n B_0 {\hat{I}_z} - \gamma_n  B_{\text{ac}} \cos(2\pi \nu t) {\hat{I}_y}.
\end{align}
According to the Heisenberg equation of motion, we can obtain the dynamical equation for the nuclear spin angular momentum as
\begin{align}
\frac{d\hat{I}_x}{dt} &= \gamma_n B_0 \hat{I}_y - \gamma_n B_{\text{ac}} \cos(2\pi \nu t) \hat{I}_z,\\
\frac{d\hat{I}_y}{dt} &= -\gamma_n B_0 \hat{I}_x,\\
\frac{d\hat{I}_z}{dt} &= \gamma_n B_{\text{ac}} \cos(2\pi \nu t) \hat{I}_x.
\end{align}
The polarization of $^{129}$Xe atoms can be defined as
\begin{equation}
\mathbf{P} = \frac{\langle \hat{\mathbf{I}} \rangle}{v_0},
\end{equation}
where $v_0^{-1}$ is the  {atomic} number density.
 {The dynamics of the three polarization components can be described by the Bloch equations as} \cite{Levitt2013}
\begin{align}
\frac{dP_x}{dt} &= \gamma_n B_0 P_y - \gamma_n B_{\text{ac}} \cos(2\pi \nu t) P_z,\\
\frac{dP_y}{dt} &= -\gamma_n B_0 P_x,\\
\frac{dP_z}{dt} &= \gamma_n B_{\text{ac}} \cos(2\pi \nu t) P_x.
\end{align}
{Generally speaking, any quantum system inevitably suffers from the interaction with its environment. On account of a Markovian noise, t}he dynamics of $^{129}$Xe spins can be described with the Bloch equations
\begin{align}
  \frac{dP_x}{dt} &= \gamma_n B_0 P_y - \gamma_n B_{\text{ac}} \cos(2\pi \nu t) P_z - \frac{1}{T_2} P_x,\\
  \frac{dP_y}{dt} &= -\gamma_n B_0 P_x - \frac{1}{T_2} P_y,\\
  \frac{dP_z}{dt} &= \gamma_n B_{\text{ac}} \cos(2\pi \nu t) P_x - \frac{1}{T_1} P_z,
\end{align}
{where $T_1$ ($T_2$) is the longitudinal (transverse) relaxation time of $^{129}$Xe spin.}
We consider the response of polarized $^{129}$Xe spins to a transverse oscillating magnetic field. {Considering a Gaussian noise, t}he dynamics of $^{129}$Xe spins can be described by the Bloch equation as \cite{breuer2002theory}
\begin{align}
  \frac{dP_x}{dt} &= \gamma_n B_0 P_y - \gamma_n B_{\text{ac}} \cos(2\pi \nu t) P_z - \frac{t}{T_2^2} P_x,\\
  \frac{dP_y}{dt} &= -\gamma_n B_0 P_x - \frac{t}{T_2^2} P_y,\\
  \frac{dP_z}{dt} &= \gamma_n B_{\text{ac}} \cos(2\pi \nu t) P_x - \frac{t}{T_1^2} P_z.
\end{align}

To analyze the system more conveniently, we transform it to the rotating frame defined by
\begin{equation}
\hat{U} = e^{i 2\pi \nu t {\hat{I}_z}}.
\end{equation}
In the rotating frame, the effective Hamiltonian
\begin{equation}
\hat{\tilde{H}} = \hat{U}^\dagger \hat{H} \hat{U} - i \hat{U}^\dagger \frac{d}{dt} \hat{U}
\end{equation}
can be simplified as
\begin{equation}
\hat{\tilde{H}} \approx \Delta {\hat{I}_z} - \frac{1}{2} \gamma_n B_{\text{ac}} {\hat{I}_y},
\end{equation}
where $\Delta=2\pi(\nu-\nu_0)$ is the detuning, and we have dropped the fast-oscillating terms by the rotating-wave approximation \cite{breuer2002theory,Aiqing2010PRA}.
Thus, the Bloch equations of $^{129}$Xe spins in the rotating frame read
\begin{align}
&\dot{\tilde{P}}_x=\frac{\gamma_{n}B_{\mathrm{ac}}}{2}\tilde{P}_{z}+\Delta\tilde{P}_{y}-\frac{t}{T_{2}^2}\tilde{P}_{x},\label{eq:BlochX}\\
&\dot{\tilde{P}}_y=-\Delta\tilde{P}_{x}-\frac{t}{T_{2}^2}\tilde{P}_{y},\label{eq:BlochY}\\
&\dot{\tilde{P}}_z=-\frac{\gamma_{n}B_{\mathrm{ac}}}{2}\tilde{P}_{x}-\frac{t}{T_{1}^2}\tilde{P}_{z},\label{eq:BlochZ}
\end{align}
where $\tilde{P}_{\alpha}~(\alpha=x,y,z)$ are the polarization of $^{129}$Xe spins in the rotating frame.
According to Eq.~(\ref{eq:Larmor}), we can define the effective magnetic field in the rotating frame as $\tilde{\mathbf{B}}=(B_{\mathrm{ac}}/2)\hat{y}-(\Delta/\gamma_n)\hat{z}$.

When the Rb pump light is off, the 
effective magnetic-field gradient induced by the Rb polarization is greatly suppressed, resulting in $T_2$ being close to $T_1$ {\cite{Su2024PRL}}. As a result, we assume {$T_2 \approx T_1=T$}  and thus we have
\begin{align}\dot{\tilde{P}}_x&=\frac{\gamma_{n}B_{\mathrm{ac}}}{2}\tilde{P}_{z}+\Delta\tilde{P}_{y}-\frac{t }{T^{2}}\tilde{P}_{x},\label{eq:Px}\\
\dot{\tilde{P}}_y&=-\Delta\tilde{P}_{x}-\frac{t }{T^{2}}\tilde{P}_{y},\label{eq:Py}\\
\dot{\tilde{P}}_z&=-\frac{\gamma_{n}B_{\mathrm{ac}}}{2}\tilde{P}_{x}-\frac{t }{T^{2}}\tilde{P}_{z}.
\label{eq:Pz}
\end{align}

\section{Linear response}
\label{sec:linear}

In practical applications, precise measurement of weak magnetic fields is often of great interest, such as those associated with precision medicine, deep-sea exploration and cardiac activity, etc {\cite{Jiang2024PNAS}}.
When $B_{\rm{ac}}$ is weak enough, we can
ignore the first term in Eq.~(\ref{eq:Pz}). In this case, the time evolution of $\tilde{P}_z$
is decoupled with $\tilde{P}_x$, $\tilde{P}_y$ and $B_{\rm{ac}}$. And $\tilde{P}_x$ exhibits a linear dependence on $B_{\rm{ac}}$.
Under the weak-field approximation, Eqs.~(\ref{eq:Px})-(\ref{eq:Pz}) can be simplified as
\begin{align}
\dot{\tilde{P}}_x&=\frac{\gamma_{n}B_{\mathrm{ac}}}{2}\tilde{P}_{z}+\Delta\tilde{P}_{y}-\frac{t}{T^2}\tilde{P}_{x},\\
\dot{\tilde{P}}_y&=-\Delta\tilde{P}_{x}-\frac{t}{T^2}\tilde{P}_{y},\\
\dot{\tilde{P}}_z&=-\frac{t}{T^2}\tilde{P}_{z}.
\end{align}
When the polarization is initially along $z$-axis, i.e., $\mathbf{P}(t=0)=(0,0,P_0)$, the solutions to the above equations can be expressed as follows
\begin{align} \label{rotating frame}
\tilde{P}_x&=\frac{P_0 B_{\rm{ac}}\gamma_n}{2\Delta} e^{-\frac{t^2}{2T^2}} \sin(\Delta t),\\
\tilde{P}_y&=\frac{P_0 B_{\rm{ac}}\gamma_n}{2\Delta}e^{-\frac{t^2}{2T^2}}[\cos(\Delta t)-1],\\
\tilde{P}_z&= P_0e^{-\frac{t^2}{2T^2}}.
\end{align}
Since the {relation} between the polarization in the rotating frame and the laboratory frame is
\begin{align}
P_{x}(t) &=\tilde{P}_x(t)\cos(\omega t)-\tilde{P}_y(t)\sin(\omega t), \\
P_y(t) &=\tilde{P}_x(t)\sin(\omega t)+\tilde{P}_y(t)\cos(\omega t), \\
P_z(t) &=\tilde{P}_z(t),
\end{align}
we can obtain the polarization in the laboratory frame as
\begin{align}
P_x&= \frac{P_0 B_{\rm{ac}}\gamma_n}{2\Delta} e^{-\frac{t^2}{2T^2}}\left[\sin(\Delta t-\omega t)+\sin(\omega t)\right],
\\P_y&=\frac{P_0 B_{\rm{ac}}\gamma_n}{2\Delta} e^{-\frac{t^2}{2T^2}}\left[\cos(\Delta t-\omega t)-\cos(\omega t)\right],
\\P_z&=P_0e^{-\frac{t^2}{2T^2}}.
\end{align}
Thus, the magnitude of the transverse polarization can be expressed as
\begin{equation}
P_\perp=\frac{P_0 B_{\rm{ac}}\gamma_n}{\Delta} e^{-\frac{t^2}{2T^2}}\sin(\frac{\Delta t}{2}).
\end{equation}
 According to $B_{\mathrm{eff}}=8\pi\kappa_0M_0P_\perp/3$,  where $\kappa_0=540$ denotes the Fermi-contact enhancement factor between $^{129}$Xe and $^{87}$Rb \cite{Walker1997,Jiang2024PNAS}{, t}he amplitude of the transverse effective field is
\begin{equation}
B_{\mathrm{eff}}=\frac{8\pi}{3}\kappa_0M_0\frac{P_0 B_{\rm{ac}}\gamma_n}{\Delta} e^{-\frac{t^2}{2T^2}}{\sin(\frac{\Delta t}{2})}.
\end{equation}
And thus the transverse field is amplified by a factor
\begin{equation}
\Pi\equiv\frac{|\mathbf{B}_{\mathrm{eff}}|}{|\mathbf{B}_{\mathrm{ac}}|} = \frac{2\lambda M_n\gamma_n}{\Delta}\sin\left(\frac{\Delta t}{2}\right)\mathrm{e}^{-\frac {t^2}{2 T^2}},
\end{equation}
where $M_n=M_0P_0$ is the nuclear magnetization of $^{129}$Xe, $\lambda=4\pi\kappa_0/3$.
The optimal time {$t_{\mathrm{opt}}$} can be obtained by calculating $\partial\Pi/\partial t |_{t=t_{\mathrm{opt}}}=0$, and thus yields $\tan\left(\Delta t_{\mathrm{opt}}/2\right)=\Delta T^2/{2t_{\mathrm{opt}}}.$
Now we expand $\tan(\Delta t_{\mathrm{opt}}/2)$ to the third order of $\Delta t_{\mathrm{opt}}$ and obtain the following expression $\left(\Delta t_{\mathrm{opt}}/2\right)+\left(\Delta t_{\mathrm{opt}}/2\right)^3/3=\Delta T^2/{2t_{\mathrm{opt}}}$.
The time required to reach the maximum transverse polarization is determined by
\begin{align}
t_{\rm{opt}} = \sqrt{\frac{6}{\Delta^2}\left(-1+\sqrt{1+\frac 1 3 \Delta^2T^2}\right)}.
\end{align}
%\begin{equation}
%t_{\rm{opt}} = T\sqrt{1\pm\sqrt{1-\frac 1 3 \Delta^2T^2}}.
%\end{equation}
Again by Taylor expansion, we could obtain
\begin{align}\label{eq:topt}
t_{\rm{opt}} \simeq T\left(1-\frac{\Delta^2T^2}{24}\right).
\end{align}
Based on this finding, it is shown that the optimal time becomes longer as $T$ increases and $\Delta$ decreases, 
which will be discussed in detail in the next section.

When the oscillation frequency of the external magnetic field matches the $^{129}$Xe Larmor frequency, i.e., $\Delta=0$, the time required to reach the maximum transverse polarization is $t_{\rm{opt}}=T$. {It should be noted that the optimal time should be smaller than $T$ in the non-resonance case, i.e., $\Delta \neq0$.} The optimal amplification factors for the Markovian and QZE cases are $\Pi=2\lambda M_n\gamma_n T \rm{e}^{-1}$ {\cite{Jiang2024PNAS}} and $\Pi=2\lambda M_n\gamma_n T \rm{e}^{-1/2}$, respectively. Therefore, {the} QZE could enhance the amplification of the weak magnetic field by a factor $\sqrt{\rm{e}}$.

\section{Nonlinear response}
\label{sec:nonlinear}

\begin{figure}[htbp]
\centering
\includegraphics[width=\columnwidth]{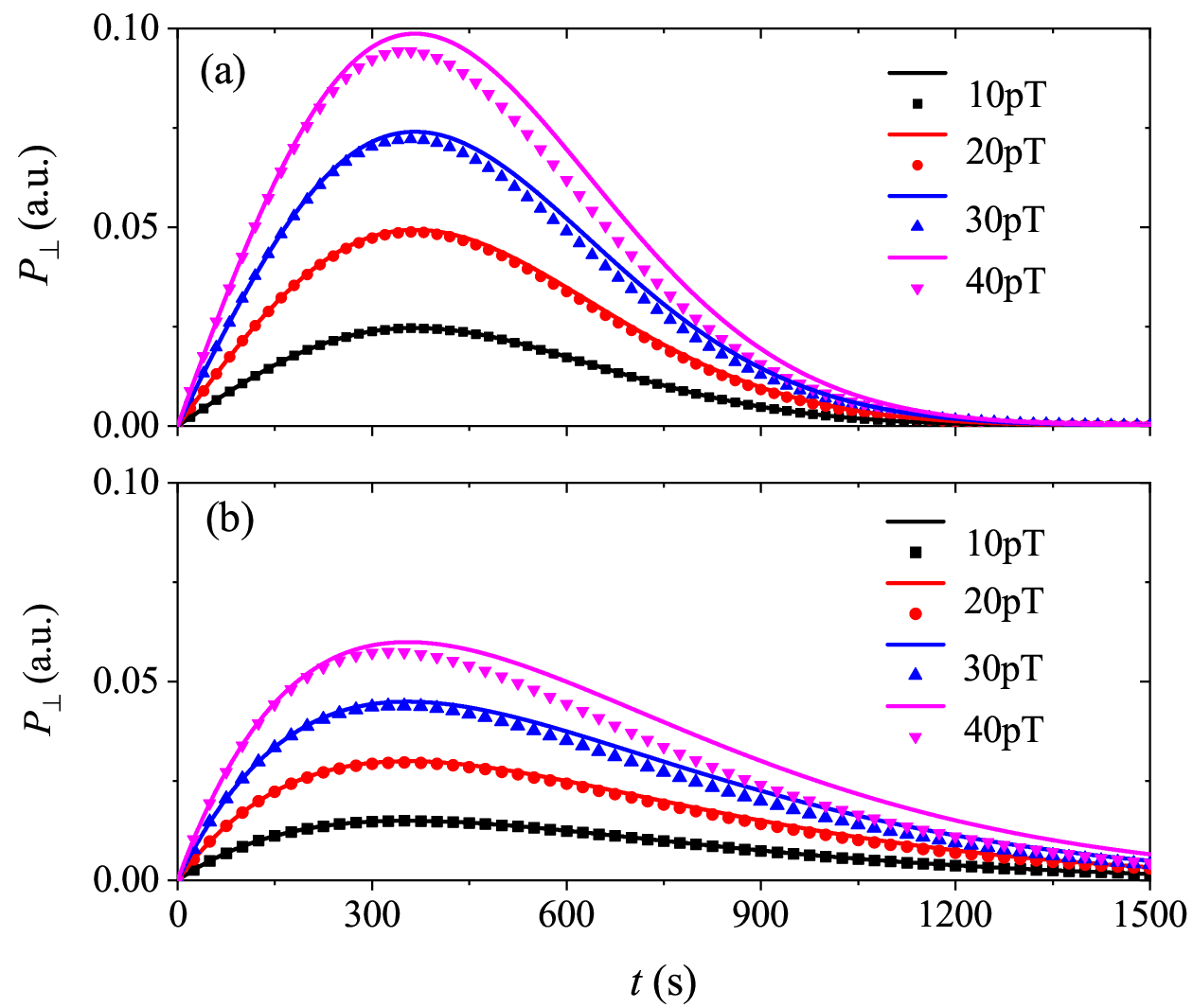}
\caption{Time evolution of the transverse polarization $P_\perp$ vs $B_{\rm{ac}}$ for (a) the Gaussian and (b) the Markovian noise, respectively. Solid lines represent the profiles under the linear-response approximation while symbols represent the profiles by the numerically-exact solution. {The parameters $T=380$~s and $\Delta=2.5$~mHz are used.}\label{fig:response}}
\end{figure}

In the following, we will obtain the response of the $^{129}$Xe spins without the weak-field approximation but by solving Eqs.~(\ref{eq:Px})-(\ref{eq:Pz}) numerically. Figure~\ref{fig:response} shows $P_\perp$ against time in the QZE and Markovian case \cite{Jiang2024PNAS},  respectively. In order to verify the validity of the weak-field approximation, we compare the linear response with the nonlinear response under different amplitudes of the magnetic field.
The approximated and the exact profiles almost overlap with each other when the measured field is 10~pT. However, as the magnetic field increases, their {difference} becomes larger and larger, which suggests that the approximation works well in the case of weak fields. Most importantly, it is clearly observed that the response in the former case is significantly larger than that in the latter, indicating that the magnetic field can be amplified more effectively by the QZE.

\begin{figure}[htbp]
\centering
\includegraphics[width=\columnwidth]
{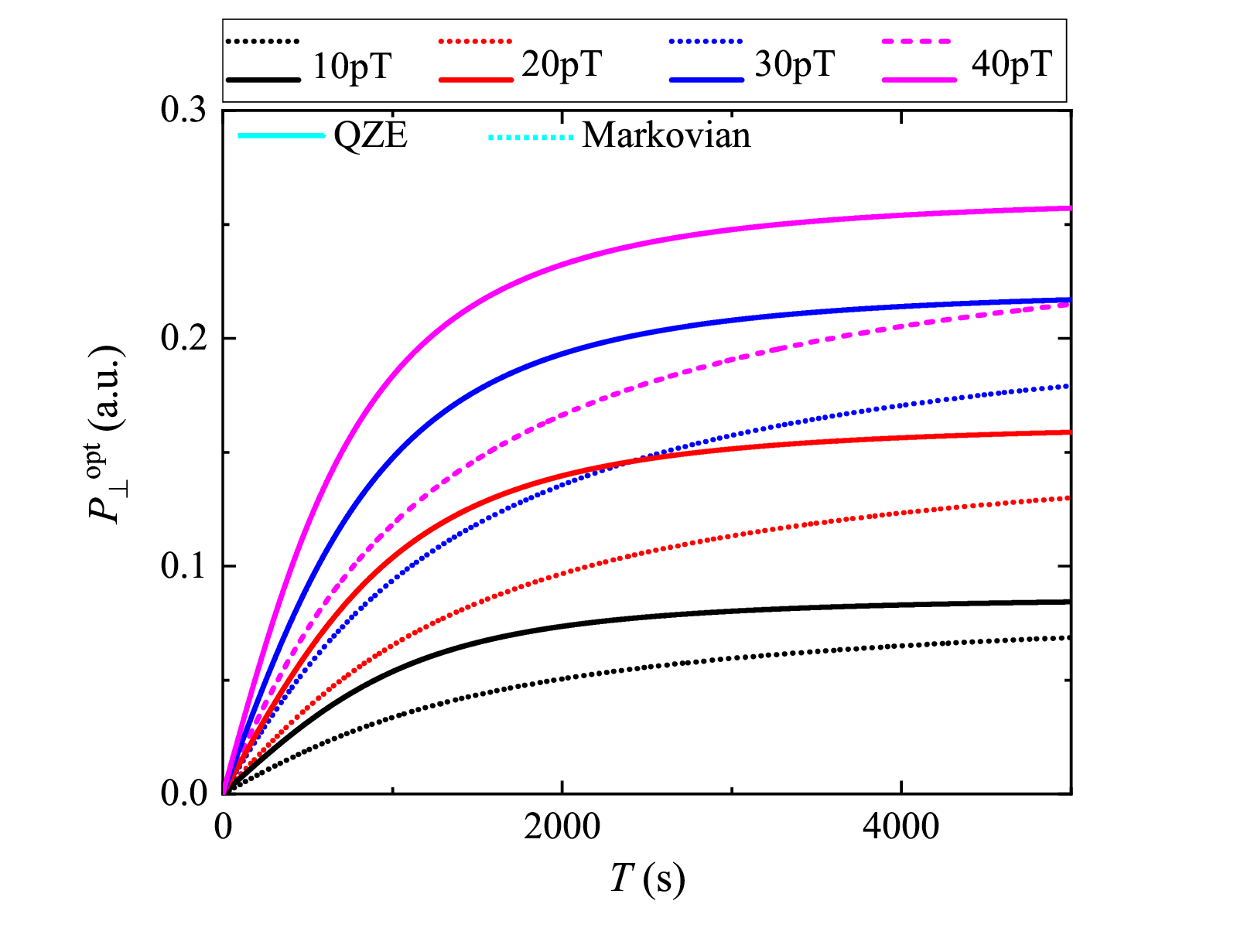}
\caption{The optimal transverse polarization {$P_{\perp}^{\rm opt}$} as a function of coherence time $T$ for different values of $B_{\rm{ac}}$ is presented for both Markovian dynamics (dotted lines) and the QZE (solid lines), respectively. The other parameters are the same as those used in Fig.~\ref{fig:response}.}\label{fig:p-T}
\end{figure}

Figure~\ref{fig:p-T} compares the optimal response {$P_{\perp}^{\rm opt}$} in the Markovian case and QZE regime over a wide range of coherence times. It is shown that the optimal responses in both cases exhibit the same trend, gradually increasing with $T$. Furthermore, the response increases with the magnitude of the magnetic field for a given coherence time $T$. Interestingly, it is evident that the responses in the QZE consistently exceed those in the Markovian case.

\begin{figure}[htbp]
\centering
\includegraphics[width=\columnwidth]
{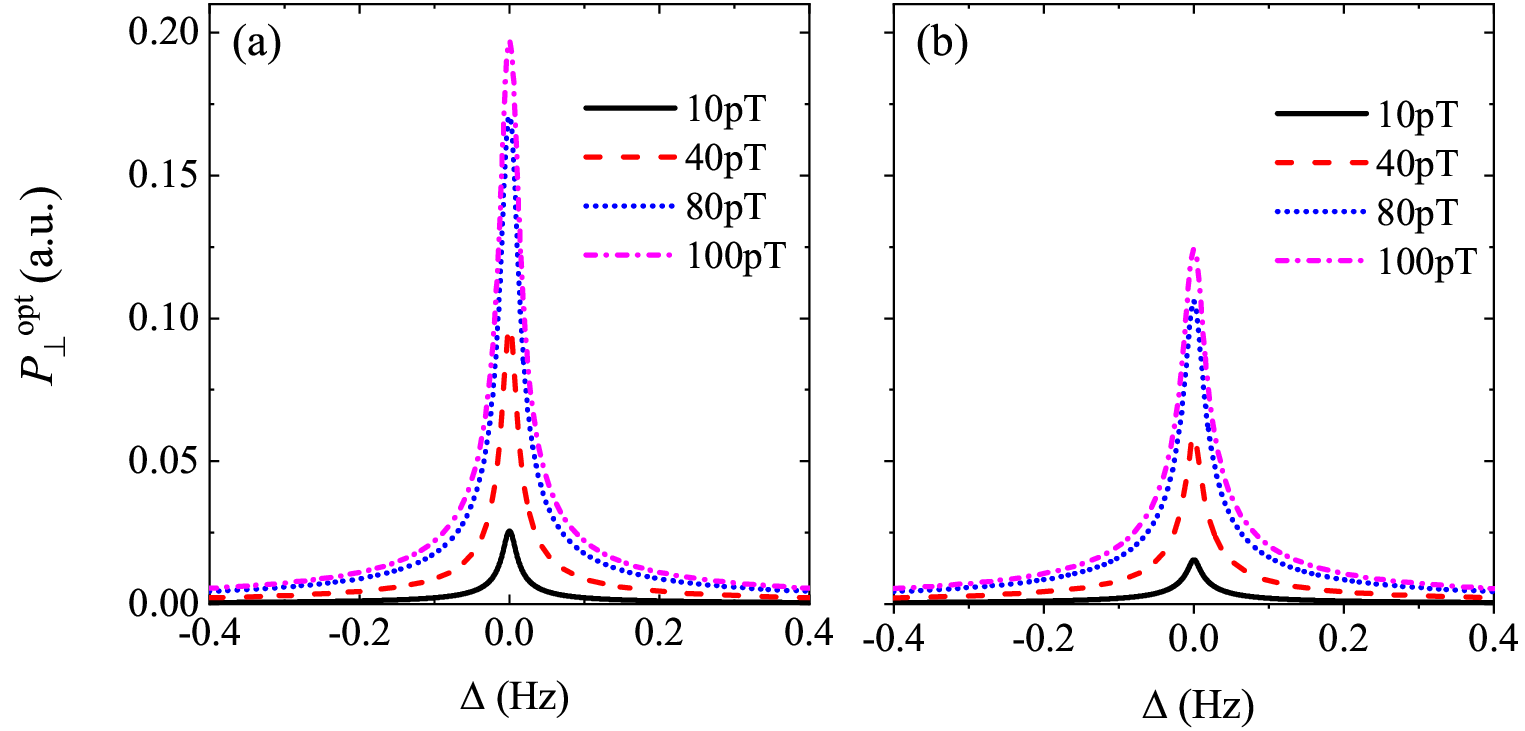}
\caption{The optimal transverse polarization {$P_{\perp}^{\rm opt}$} in (a) the QZE and (b) the Markovian cases as a function of detuning $\Delta$ for different $B_{\rm{ac}}$s. The other parameters are the same as those used in Fig.~\ref{fig:response}.}\label{fig:p-delta}
\end{figure}

In order to evaluate the effect of the detuning $\Delta$ on the amplification of the magnetic field, as shown in Fig.~\ref{fig:p-delta}, we compare the optimal response{s} under the QZE with {those} in the Markovian case for different $B_{\rm{ac}}$s by solving Eqs.~(\ref{eq:Px})-(\ref{eq:Pz}) numerically. As illustrated in Fig.~\ref{fig:p-delta}(a), the optimal response decreases monotonically with increasing detuning for a fixed value of $B_{\rm{ac}}$. Therefore, in order to achieve a larger amplification of the weak magnetic field, the resonance between the driving field and Larmor frequency is required. Furthermore, the optimal response shows a gradual increase as the magnetic field strength is enhanced for a constant detuning. Since we employ the QZE to amplify the weak magnetic field, we further investigate the optimal responses in the Markovian noise in Fig.~\ref{fig:p-delta}(b). For all $B_\textrm{ac}$s, the optimal response is always smaller than the counterparts in the QZE.

%The observed response behavior of in-the-dark $^{129}$Xe spins are in good agreement with our analysis results described in Eq. \eqref{topt}.
%It reveals that, with the enlargement of the detuning, not only are the responses suppressed in both cases, but the difference between them becomes also smaller.

\begin{figure}[htbp]
\centering
\includegraphics[width=\columnwidth]
{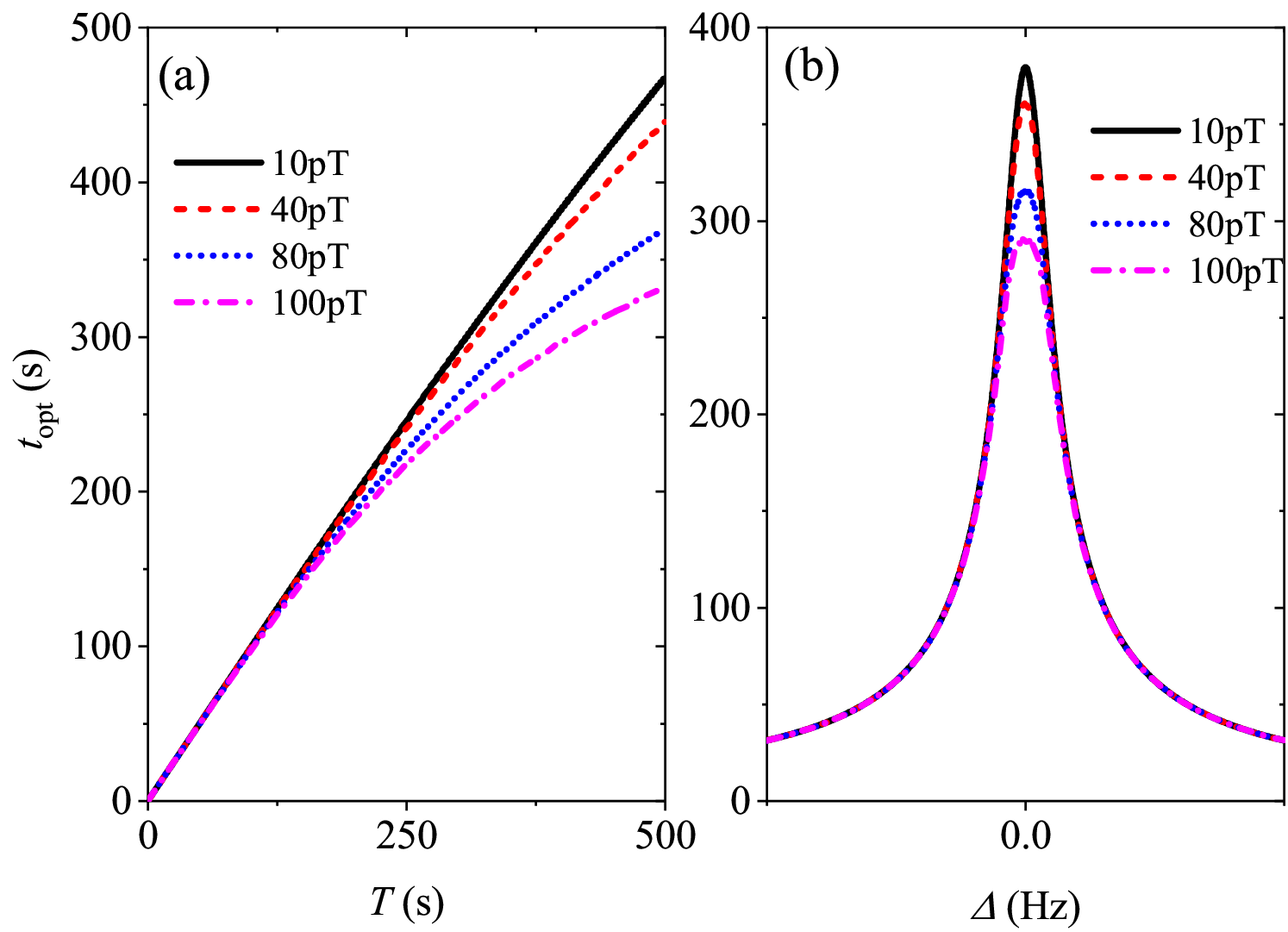}
\caption {Dependence of the optimal time $t_{\textrm{opt}}$ on (a) the relaxation time $T$ with $\Delta=2.5$~mHz and (b) the detuning $\Delta$ with $T=380$~s for the QZE in different $B_{\rm{ac}}$s.}\label{fig:tmax-T-Delta}
\end{figure}

In Fig.~\ref{fig:tmax-T-Delta}, we present the optimal time $t_{\textrm{opt}}$ to perform the measurement as a function of the relaxation time $T$ and detuning $\Delta$ for the QZE, respectively.
It is demonstrated that $t_{\textrm{opt}}$ is proportional to $T$ for $B_{\rm{ac}}=10~$pT, which is consistent with Eq.~(\ref{eq:topt}). However, as $B_{\rm{ac}}$ increases, the time required to achieve the optimal response decreases for a given $T$. On the other hand, when talking about the dependence of the optimal measurement time on $\Delta$, it turns out to be much complicated. It requires the longest time to achieve the optimal response for the resonance case, while {$t_{\textrm{opt}}$} decreases as the detuning is enlarged, which is predicted by Eq.~(\ref{eq:topt}). Notably, for different magnetic fields, all curves coincide with each other in the large-detuning limit.
These findings suggest that maximizing the relaxation time while minimizing the detuning is crucial for expediting the amplification of the weak magnetic field. Thus, careful control of these parameters is essential to enhance the efficiency and precision of the measurement process.

\begin{figure}[htbp]
\centering
\includegraphics[width=\columnwidth]
{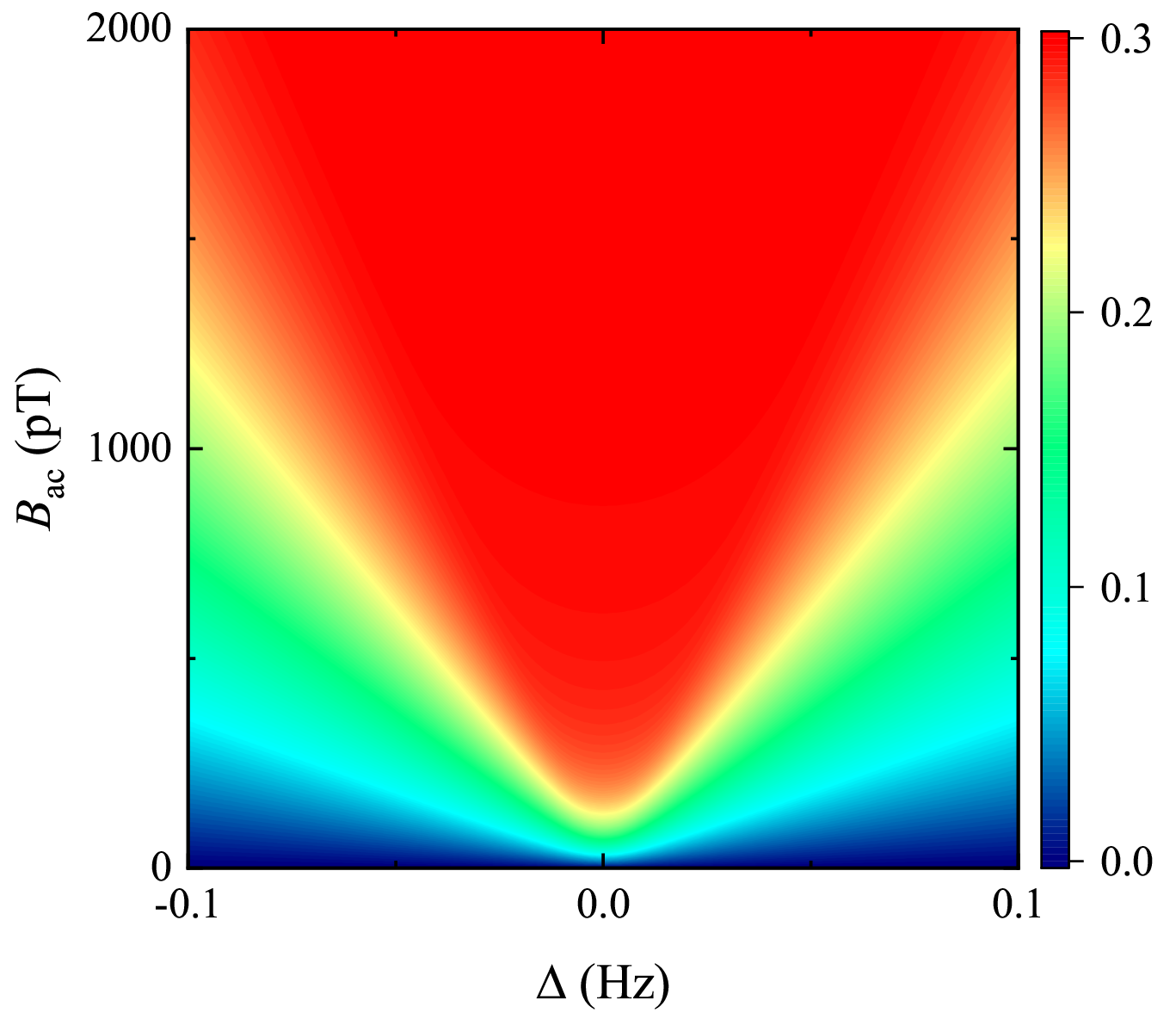}
\caption{Phase diagram of the optimal response $P_{\perp}^{\rm opt}$ vs $B_{\rm{ac}}$ and $\Delta$ for $T=380$~s. }\label{fig:B-Delta}
\end{figure}

Since in Fig.~\ref{fig:p-delta}(a), {the optimal responses increase as the magnetic field is enlarged,} it may be quite natural to ask whether there exists a bound for the optimal response. As depicted in Fig.~\ref{fig:B-Delta},
we explore the optimal response for different sets of ($B_{\rm{ac}}, \Delta$). When the system is at resonance, the optimal response will quickly be saturated as the magnetic field is enlarged. As the detuning increases, the smallest magnetic field to saturate optimal response becomes lager and larger.

%Our analysis reveals that the smaller the $\Delta$, the larger the optimal response under relatively weak magnetic fields. Conversely, the optimal measurement value becomes independent of $\Delta$ when the field strength exceeds approximately 180 pT.
%Besides, the amplification of magnetic field is better in the resonant situation than in the far-off-resonant situation when $B_{\rm{ac}}$ is weak enough.

\begin{figure}[htbp]
\centering
\includegraphics[width=\columnwidth]
{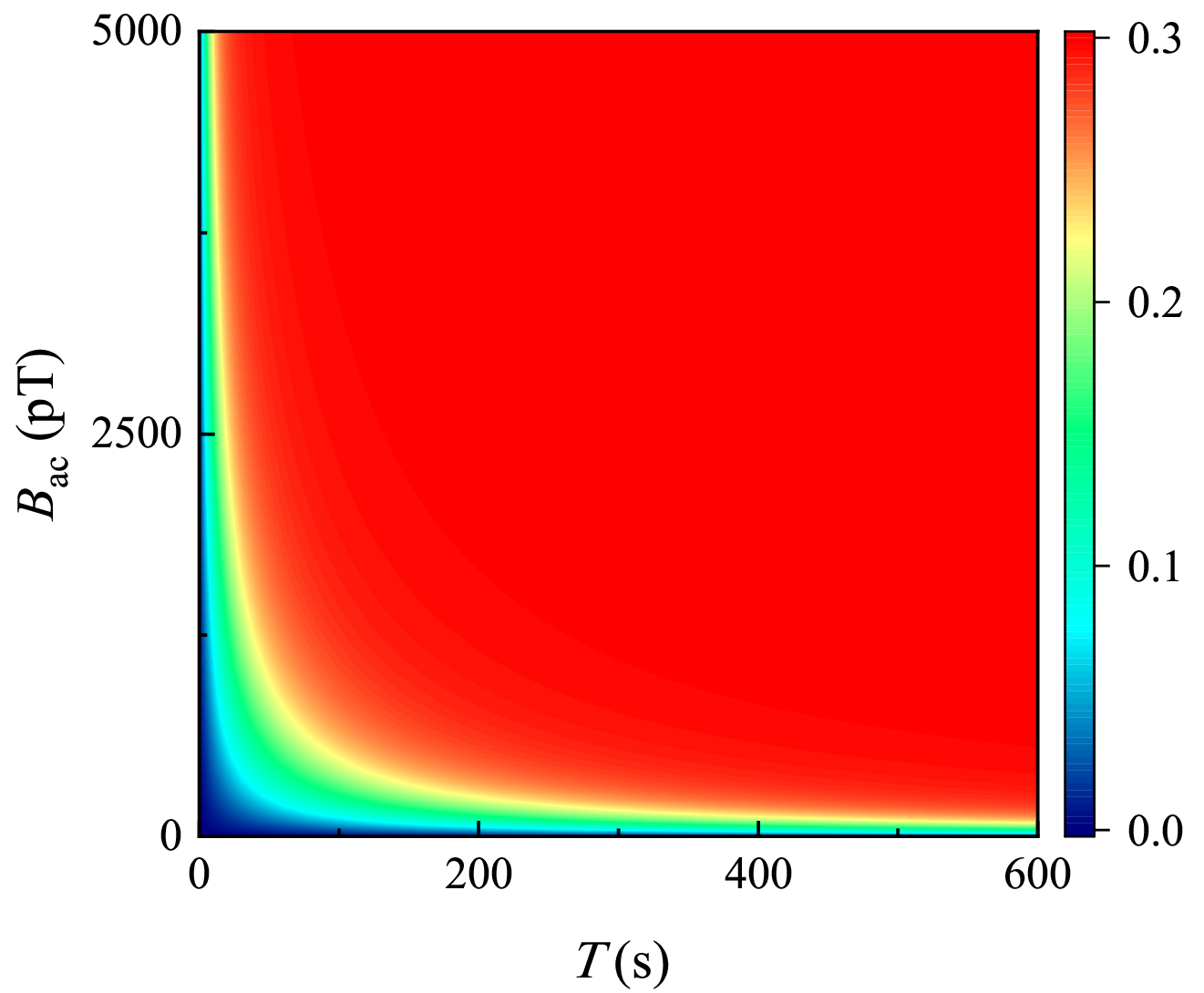}
\caption{Phase diagram of the optimal response $P_{\perp}^{\rm opt}$ vs $B_{\rm{ac}}$ and $T$ for $\Delta=2.5$~mHz. }\label{fig:B-T}
\end{figure}

To gain further insight into the effects of the amplitude of the oscillating field $B_{\rm{ac}}$ and the relaxation time $T$ on the optimal response {$P_{\perp}^{\rm opt}$}, we focus on the near-resonance case and present the phase diagram of the optimal response for a large range of $T$ and $B_{\rm{ac}}$ in Fig.~\ref{fig:B-T}. It is shown that the optimal response can be achieved across the majority of the parameter space $(B_{\rm{ac}},T)$, which provides convenience for measuring magnetic fields of various amplitudes. However, it should be pointed out that the relaxation time $T$ should be sufficiently large, e.g. $T>60$~s, in order to attain the maximum optimal response larger than 0.28.

\begin{figure}[htbp]
\centering
\includegraphics[width=\columnwidth]
{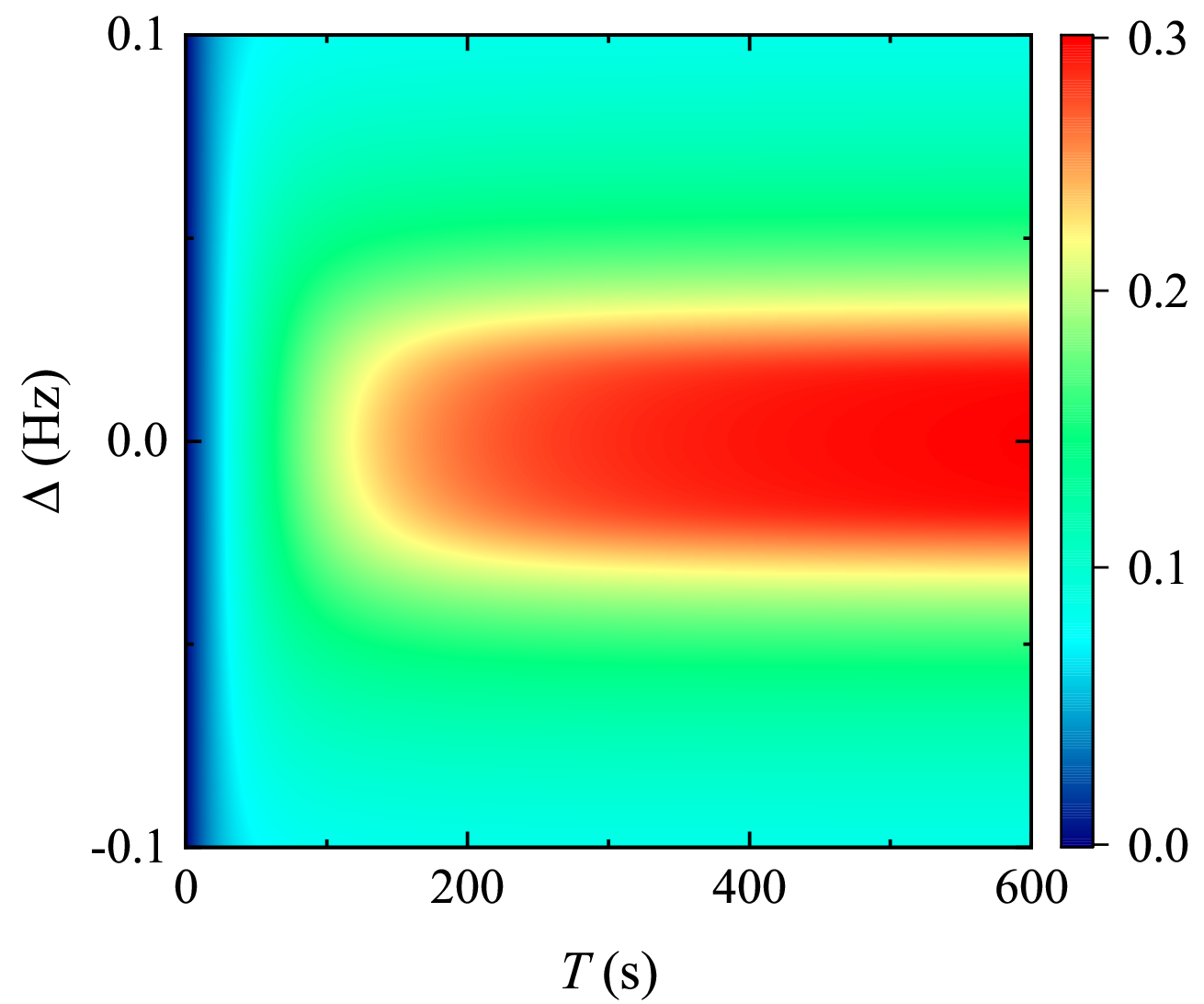}
\caption{Phase diagram of the optimal response {$P_{\perp}^{\rm opt}$} vs $\Delta$ and $T$ for $B_{\rm{ac}}=400$~pT. The other parameters are the same as those in Fig.~\ref{fig:response}.}\label{fig:Delta-T}
\end{figure}

In Fig.~\ref{fig:Delta-T}, we explore the optimal response in the parameter space ($\Delta,T$) for $B_{\rm{ac}}=400$ pT. The results indicate that the optimal response is confined to a narrow parameter region, characterized by a large relaxation time $T$ and a small detuning $\Delta$, as depicted by the red region. 

\begin{figure}[htbp]
\centering
\includegraphics[width=\columnwidth]
{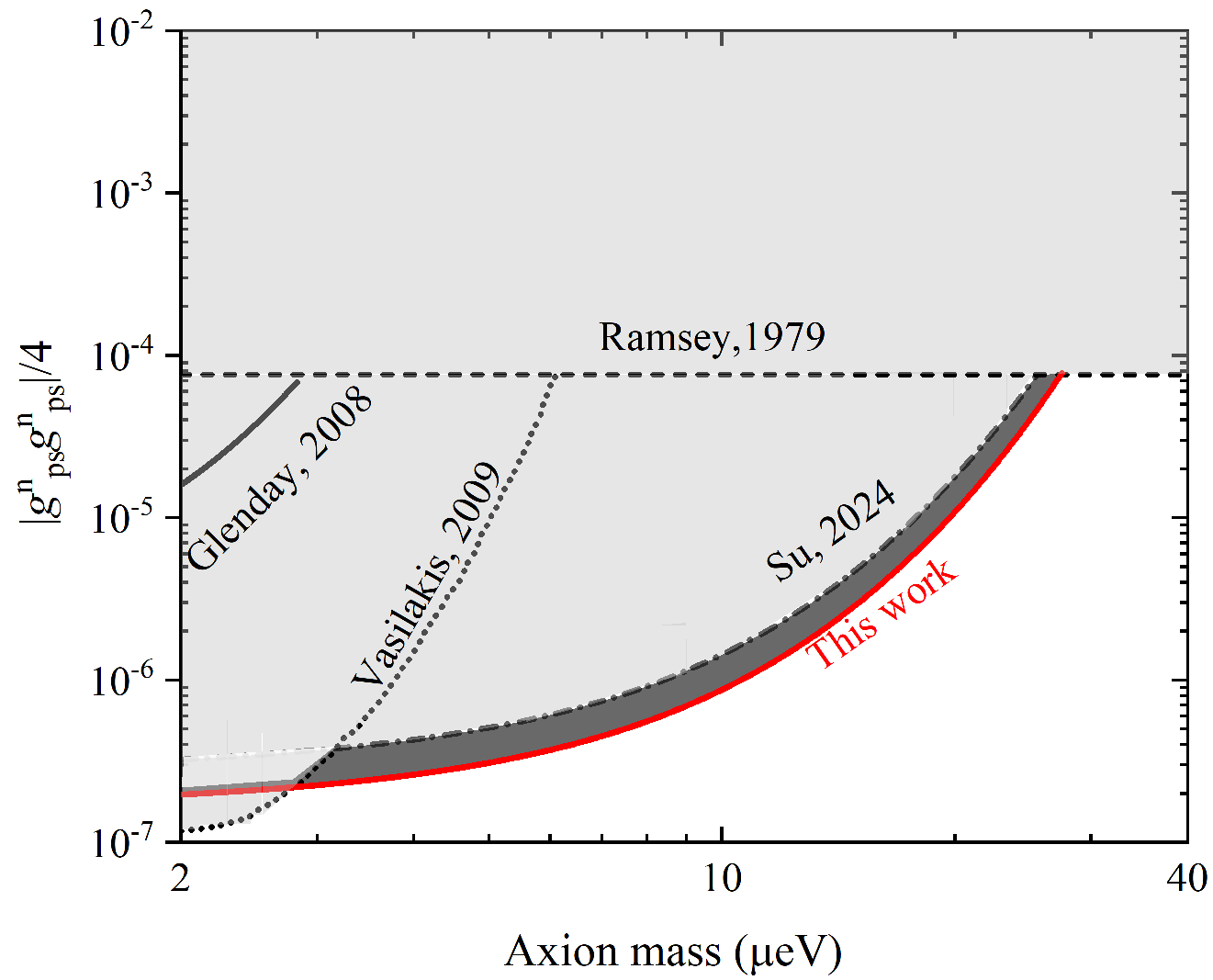}
\caption{Constraint{s} to the coupling{-}constant{s} product $|g^n_{ps}g^n_{ps}|/4$ within the axion window. The black curve{s} represent the experimental limits on the neutron-neutron coupling from the previous experiments of Ramsey \cite{Ramsey1979PhysicaA}, Glenday \textit{et al.} \cite{Glenday2008PRL}, Vasilakis \textit{et al.} \cite{Vasilakis2009PRL} and Su \textit{et al.} \cite{Su2024PRL} as a function of the axion mass.
The red line represents the new constraints on axions or {ALPs} obtained from our theory, which establishes an improved bound by a factor about $e^{1/2}$ due to the QZE.}\label{fig:gg-m}
\end{figure}

According to Refs.~\cite{Moody1984PRD,Bogdan2006JHEP,Fadeev2019PRA}, the interaction between the polarized neutrons is
\begin{align}
V_{\rm ps-ps}=&\frac{g_{\rm ps}^{n}g_{\rm ps}^{n}}{16\pi m_n^2}\left[\hat{\mathbf{\sigma}}_{\rm so}\cdot\hat{\mathbf{\sigma}}_{\rm se}\left(\frac{m_a}{r^2}+\frac{1}{r^3}\right)\right.\nonumber\\
&\left.-(\hat{\mathbf{\sigma}}_{\rm so}\cdot\hat{r})(\hat{\mathbf{\sigma}}_{\rm se}\cdot\hat{r})\left(\frac{m_a^2}{r}+\frac{3m_a}{r^2}+\frac{3}{r^3}\right)\right],
\end{align}
where $\hbar=c=1$,  $g_{\rm ps}^{n}$ is the pseudoscalar coupling constant of the neutron, 
$m_n$ ($m_a$) is the mass of the neutron (axion), $\hat{\mathbf{\sigma}}_{\rm so}$ ($\hat{\mathbf{\sigma}}_{\rm se}$) is the spin operator in the source (sensor) cell, $\vec{r}=r\hat{r}$ is the distance vector between two interacting neutrons. {$B_{\rm ac}\cos(2\pi\nu t)\hat{\mathbf{\sigma}}_{\rm so}=-V_{\rm ps-ps}/\nu_{\rm Xe}$} is the pseudomagnetic field experienced by the nuclear spins in the sensor cell, where {$\nu$} is the Larmor frequency of the nuclear spin in the sensor cell. Following the same procedure in Ref.~\cite{Su2024PRL}, 
we can effectively estimate $|g^n_{ps}g^n_{ps}|/4$.
Figure~\ref{fig:gg-m} shows the obtained constraints on $|g^n_{ps}g^n_{ps}|/4$ set by this work together with the limits from the previous experimental searches \cite{Ramsey1979PhysicaA,Glenday2008PRL,Vasilakis2009PRL,Su2024PRL}. The excluded values of the coupling-constants product of the previous results are presented as the light-gray areas and the result of this work is presented as the dark-gray area. The first constraint of exotic spin-spin interactions was derived by Ramsey \cite{Ramsey1979PhysicaA} and presented as the black dashed line. The black solid and dotted lines respectively represents the constraint{s} on $|g^n_{ps}g^n_{ps}|/4$ placed by the maser \cite{Glenday2008PRL} and the comagnetometer \cite{Vasilakis2009PRL} experiments. Based on the recent work \cite{Su2024PRL}, we set the most stringent constraint on $|g^n_{ps}g^n_{ps}|/4$ by utilizing the QZE from a theoretical perspective and presented as the red solid line, part of which reaches into the unexplored parameter space within the axion window. For the mass range from 3.2 to 24.3~$\mu$eV, we improve the previous constraint by a factor about $e^{1/2}$ within the axion window. The major improvement in the constraint comes from the QZE. In addition, our theoretical approach can also be utilized to search for other exotic spin-dependent interactions, such as those between the polarized electrons, and between a polarized electron and an unpolarized nucleon. Since it has not yet been implemented in the experiments, we do not only provide the optimal measurement parameters but also obtain optimal amplification, i.e., about 8900, and the corresponding time for the practical experiments, which can effectively guide related experiment in this work. We hope that the experiment could be realized in this overlapping spin ensemble by the QZE in the near future, contributing to the detection of exotic interactions and ALPs.

%In a word, We determine that achieving a stronger response requires operating under resonant conditions whenever possible by an analysis of the parameter space response. Additionally, maximizing the relaxation time and magnetic field strength further enhances the response. These findings offer valuable insights for the design of related experiments.

\section{Conclusion}
\label{sec:Conclusion}

In this paper, we propose a noble-gas spin evolution in the dark and amplifying the measurement of the weak magnetic field by the QZE. Recently, there has been significant progress in amplifying the weak magnetic field with mixtures of nuclear-spin-polarized noble gases and vapors of spin-polarized alkali-metal atoms, such as $^{129}$Xe-$^{87}$Rb \cite{Su2021science,Jiang2021NP,Jiang2024PNAS}. The experimental process generally includes three necessary steps. First of all, we polarize alkali-metal spins with optical pumping and then noble-gas nuclear spins are polarized through spin-exchange collisions with polarized alkali-metal atoms spins. Secondly, the transverse magnetization generated by these nuclear spins produces effective
fields on alkali-metal spins, which leads to magnetic amplification process. Finally, we measure the effective magnetic fields and calibrate the parameters of noble gases' spins with the magnetometer of alkali-metal atoms.
In general, the dynamics of any open quantum system is initialized by a Gaussian decay \cite{Nakazato1996IJMPB}, where the QZE occurs. It is followed by the Markovian dynamics and finalized by an power-law decay. However, the QZE is not restricted to the open quantum systems and can be generalized to closed quantum systems, which is an intrinsic effect due to the unitarity in quantum mechanics. As experimentally demonstrated in Ref.~\cite{Nagels1997PRL}, the decoherence time $T$ can be effectively prolonged by decreasing the pressure of the overlapping-spin ensemble.
This technique can also be applied to various alkali-metal atoms and noble gases, including
K-$^{129}$Xe, K-$^3$He and $^{87}$Rb-$^{21}$Ne, etc. In the experiments, the two main factors that affect the coherence time of the noble gas are the effective magnetic-field gradient from the polarization of alkali-metal atoms and the magnetic-field gradient from the applied field \cite{Jiang2024PNAS}. Therefore, decreasing the pressure of the overlapping-spin ensemble can not affect coherence time of the noble gas. If the QZE is applied to this technique, this series of experiments will be enhanced with a more significant magnetic-field amplification effect.
And thus it can be used to search for axions, dark photons and axion-mediated spin-dependent interactions, etc.
%In this system, QZE can be controlled independently by adjusting ...
Since the results of this work provide more significant magnetic amplification, we hope that our studies will stimulate experiments on establishing new constraints of dark matter and exotic interactions by this method.

In summary, we exploit the magnetic amplification considering {the} QZE through effective fields from collisions between alkali-metal atoms and noble{-}gas atoms, increasing the magnetic magnification by up to about 1.65-fold relative to the Markovian {noise}. Based on our analysis, we obtain the optimal time required to reach the maximum transverse polarization and thus the optimal response under different combinations of the parameters. Our results indicate that the amplification of measuring the weak fields can be further enhanced by the QZE as compared to the Markovian case. This indicates that our research can further improve the accuracy of weak field measurements.

\begin{acknowledgments}

This work is supported by Innovation Program for Quantum Science and Technology under Grant No.~2023ZD0300200, the National Natural Science Foundation of China under Grant No.~62461160263, Beijing Natural Science Foundation under Grant No.~1202017, and Beijing Normal University under Grant No.~2022129.

\end{acknowledgments}

%\appendix
%
%\section{Population of Atomic Excited State}
%\label{AppendixA}

%The authors declare no conflict of interest.

%\bibliographystyle{alpha}
%\bibliography{ref}
%\begin{thebibliography}{29}%

\bibliography{ref}
	
\end{document}